\RequirePackage{silence}
\documentclass[fleqn,usenatbib,useAMS]{mnras} 
\usepackage{graphicx}
\usepackage{amsmath}

\usepackage{amsfonts}
\usepackage{float}
\usepackage{bm}
\usepackage[perpage,symbol*]{footmisc}
\usepackage{ae,aecompl}
\usepackage{array}
\usepackage{soul}
\usepackage{mathtools}
\usepackage{multirow}

\newcommand{\appropto}{\mathrel{\vcenter{
			\offinterlineskip\halign{\hfil$##$\cr 
				\propto\cr\noalign{\kern2pt}\sim\cr\noalign{\kern-2pt}}}}}

\defcitealias{Speagle+14}{SP14}

\title[Star formation histories of nearby galaxies]{Constraints on the star formation histories of galaxies in the Local Cosmological Volume} 

\author[Kroupa et al.]{
	P. Kroupa,$^{1,2}$\thanks{Email: \href{mailto:pkroupa@uni-bonn.de}{pkroupa@uni-bonn.de}/\href{mailto:kroupa@sirrah.troja.mff.cuni.cz}{kroupa@sirrah.troja.mff.cuni.cz} (PK)} 
	M. Haslbauer,$^{1}$
	I. Banik,$^1$\thanks{Alexander von Humboldt Fellow}
	S.T. Nagesh,$^3$
	and J. Pflamm-Altenburg$^1$ \\
	$^1$Helmholtz-Institut f\"ur Strahlen- und Kernphysik, Universit\"at Bonn, Nussallee 14-16, 53115 Bonn, Germany \\
	$^2$Charles University in Prague, Faculty of Mathematics and Physics, Astronomical Institute, V  Hole\v{s}ovi\v{c}k\'ach 2, CZ-180 00 Praha,\\Czech Republic \\
	$^3$Argelander-Institut f\"ur Astronomie, Auf dem H\"ugel 71, 53121 Bonn, Germany
}

\pubyear{2020}
\begin{document}
\label{firstpage}
\pagerange{\pageref{firstpage}--\pageref{lastpage}}

\maketitle

\begin{abstract} The majority of galaxies with current star-formation rates (SFRs), $SFR_{\rm o} \ge 10^{-3} \, M_\odot$/yr, in the Local Cosmological Volume where observations should be reliable, have the property that their observed $SFR_{\rm o}$ is larger than their average star formation rate.  This is in tension with the evolution of galaxies described by delayed-$\tau$ models, according to which the opposite would be expected. The tension is apparent in that local galaxies imply the star formation timescale $\tau \approx 6.7$~Gyr, much longer than the $3.5-4.5$~Gyr obtained using an empirically determined main sequence at several redshifts. Using models where the SFR is a power law in time of the form $\propto \left( t - t_1\right)^\eta$ for $t_1 = 1.8$~Gyr (with no stars forming prior to $t_1$) implies that $\eta = 0.18 \pm 0.03$. This suggested near-constancy of a galaxy's SFR over time raises non-trivial problems for the evolution and formation time of galaxies, but is broadly consistent with the observed decreasing main sequence with increasing age of the Universe.  \end{abstract}

\begin{keywords} 
	galaxies: star formation -- galaxies: stellar content -- galaxies: evolution -- galaxies: formation -- Galaxy: evolution -- Galaxy: formation
\end{keywords}

\section{Introduction}

The formation and evolution of observed galaxies constitutes a cosmological boundary condition which must be fulfilled by any physically realistic theory of cosmological structure formation and evolution. A Milky Way (MW) type galaxy is generally thought to have begun forming $\approx 12$~Gyr ago \citep[i.e. at time $t_1\approx 1.8$~Gyr after the Big Bang,][]{Planck_2015}, reaching a maximum star formation rate (SFR)\footnote{SFR is the acronym used for star formation rate, while $SFR$ is the associated physical parameter.} with the SFR subsequently decreasing \citep[e.g.][]{Mor+19}.

The MW is a well-studied case, for which measurements of the phase-space, chemical, and age parameters of a large ensemble of disk stars reveal a complex star-formation history (SFH). It decreases from $SFR\approx 10-20\,M_\odot$/yr about 10~Gyr ago to a minimum with $SFR\approx 2.7\,M_\odot$/yr about $6\,$Gyr ago, reaching a local maximum of $SFR\approx 8\,M_\odot$/yr about $3.5\,$Gyr ago and decreasing to a present-day value of $SFR_{\rm o} \approx 1.3\,M_\odot$/yr \citep{Mor+19}. While the data may also be interpreted with a constant $SFR\approx 4\,M_\odot$/yr over the past $\approx 12\,$Gyr (yielding a present-day disk stellar mass of $M_* \approx 4.8 \times 10^{10}\,M_\odot$), a general trend of a decreasing SFR has also been found by \citet{Zonoozi+19} based on the shape of the galaxy-wide stellar mass function, and from an analysis of the kinematical imprint of star-formation in a clustered mode according to which the thick disk may have been formed from an elevated SFR about $10-12$~Gyr ago \citep{Kroupa02}. \citet{Ruiz-Lara+20} map the SFH from $>12$~Gyr ago until today, finding consistency with the analysis by \citet{Mor+19} and showing time-resolved evidence that it may have been modulated over the past 6~Gyr by the orbit of the Sagittarius satellite galaxy.\footnote{This raises the question of why orbital decay through dynamical friction between the putative dark matter haloes of the satellite and the MW has not yet merged the two \citep{Kroupa2015}.} In the context of Milgromian Dynamics \citep[MOND,][]{Milgrom_1983, BM84}, the SFH of the MW is likely to have been significantly affected by its encounter with Andromeda (M31) about 7-10$\,$Gyr ago \citep{Zhao+13}.\footnote{The plane or disk of satellite galaxies orbiting the MW \citep{Kroupa+05}, as confirmed by the most recent Gaia proper motion data \citep{PK20}, appears to be strongly correlated with the plane of satellite galaxies around M31 \citep{Ibata_2013} by both being polar relative to the MW and having their orbital angular momentum vectors within $45^{\rm o}\pm7^{\rm o}$ of each other \citep{Pawlowski+13}. This configuration is most likely a natural outcome of the MW-M31 encounter in MOND \citep{Banik+18, Bilek+18}.} The MW SFH is thus not likely that of an unperturbed self-regulated secularly-evolving disk galaxy. The other large galaxy in the Local Group, M31, had $SFR \approx 5\,M_\odot$/yr from 14 to 8~Gyr ago, with a dip to $SFR\approx 1\,M_\odot$/yr about 8--6~Gyr ago, a maximum near $5\,M_\odot$/yr about 5--1.5~Gyr ago and a decline to $SFR_{\rm o}\approx 0.1\,M_\odot$/yr or less (\citealt{Williams+17}, the quoted values comprise only the region of M31 mapped by the PHAT survey). It appears that the SFHs of the MW and M31 may be correlated, as the initial larger values, the minima and maxima appear to be broadly in-phase.

On the other hand, the Triangulum galaxy (M33) is a near-flocculent disk galaxy in the Local Group without a significant bulge and appears to have evolved largely without significant perturbations. Its SFR has increased significantly until the present time, with $SFR_{\rm o}\approx 0.5\,M_\odot$/yr, while about~10$\,$Gyr ago, $SFR\approx 0.04\,M_\odot$/yr with evidence for an acceleration and then minimum about 5~Gyr ago \citep{Javadi+17}. This SFH may be related to its orbit around M31 \citep{Patel+17}. The Large Magellanic Cloud (LMC) also shows an increasing SFR until today with a dip about 5~Gyr ago \citep{Meschin+14}. This SFH cannot be representative of an isolated self-regulated disk galaxy because the LMC is strongly interacting with the Small Magellanic Cloud and also with the MW.  It may be noteworthy that all four major galaxies of the Local Group discussed above share a similar depression in their SFHs roughly $5\,$Gyr ago. The evidence from the larger galaxies in the Local Group concerning the shapes of their SFHs is thus somewhat ambiguous and is mired by the Local Group members being subject to relatively strong encounters. The rather strikingly symmetrical arrangement of matter within the Local Group appears to be additionally troubling \citep{Pawlowski+13}.

In order to improve our understanding of the formation and evolution of galaxies, larger and representative samples are needed. \citet{Delgado-Serrano+10} used Sloan Digital Sky Survey (SDSS) data to show that the vast majority of galaxies with stellar masses $M_* > 1.5 \times 10^{10}\,M_\odot$ are star-forming disk galaxies in the Local Universe as well as~$6\,$Gyr ago, with elliptical galaxies comprising an unchanging $3-4\%$ of the studied ensembles. The observational study by \cite{DePropris+14} of merger rates using close pairs indicates blue galaxies to have a small merger incidence, less than that expected in standard-cosmology. \cite{DePropris+14} find ``Galaxies in these close pairs are likely to have undergone a series of previous encounters and close passes'' which poses a problem for the expected rapid merging due to dynamical friction on the large and massive dark matter haloes \citep{Kroupa2015, Oehm+17}. \cite{DePropris+14} conclude ``At face value our findings minimize the importance of major mergers and interactions for galaxy evolution and argue that most galaxy evolution takes place via internal and secular processes.'' Based on a survey of major disk galaxies within a distance of $\approx 8$~Mpc, \cite{Kormendy+10} find about half of the sample of 19 massive disk galaxies have no evidence for a classical bulge, raising the question ``How can hierarchical clustering make so many giant, pure-disk galaxies with no evidence for merger-built bulges?'' Comparing models of hierarchical galaxy formation with SDSS data, \cite{Shankar+14} conclude ``Despite the observational uncertainties, the data tend to disfavour hierarchical models characterised by strong and impulsive disc instabilities, strong gas dissipation in major mergers, short dynamical friction time-scales.'' 
In "The Impossibly Early Galaxy Problem", \cite{Steinhardt+16} note $10^{12-13}\,M_\odot$ dark matter halos to be overabundant at $z=6-8$ than expected if hierarchical assembly through mergers were to be valid. Mergers may thus be playing a lesser role in galaxy evolution than currently thought. The vastly dominating class of disk galaxies may thus be simpler than thought. Indeed, a principal component analysis of the galaxies selected with 21$\,$cm-observations implies star-forming galaxies are governed by one parameter only, making such galaxies simpler than expected \citep{Disney+08}. Disk galaxies are well-known to obey well-defined Milgromian laws \citep{Milgrom_1983,BM84}, such as the baryonic-Tully-Fisher relation \citep{McGaugh12} and the radial-acceleration relation \citep{Lelli+17} $-$ for a review, see \citep{FamaeyMcGaugh12}.

In view of this documented simplicity of galaxies, it is also important to quantify if the SFHs typically increase or decrease with time, because this question is intimately linked to the cosmological matter cycle and budget. By compiling many different surveys of star-forming galaxies with $M_* >10^9\,M_\odot$, \citet[][hereafter \citetalias{Speagle+14}]{Speagle+14} found them to form a well defined tight $SFR$ vs $M_*$ correlation, which is known as the main sequence of galaxies \citep[e.g.][]{Dave08, Renzini_2015, Popesso+19}. With increasing cosmological redshift $z$, the main sequence evolves to larger $SFR$ for a given $M_*$. The tight correlation of the main sequence across cosmic time and the apparent simplicity of galaxies can be distilled into the understanding that while simple, galaxies tend to evolve according to `delayed-$\tau$' models: After first increasing from an age, $t$, of the Universe of about $t=3\,$Gyr ($10.8\,$Gyr ago), after $t\approx 6.5-7.5\,$Gyr their SFRs decrease as their gas is consumed through star formation \citepalias[fig.~10 in][]{Speagle+14}.\footnote{The reader is referred to the discussion in \citetalias{Speagle+14} concerning the expected stochastic evolution of galaxies due to mergers as predicted in the standard dark-matter based models of structure formation, versus the more deterministic models such as through cold accretion of gas \citep{Keres+05, Dave08}. Following \citetalias{Speagle+14}, it is implicitly assumed here that ``galaxies simply follow a common, deterministic track''.} These studies suggest that the SFHs of galaxies can be described by the delayed-$\tau$ model (Eq.~\ref{eq:deltau} below) and consequently that the present-day SFR of a typical galaxy\footnote{`Typical galaxy' means a galaxy which has not suffered a major encounter within the past few dynamical times such that its SFH is approximately that of an unperturbed galaxy.}, $SFR_{\rm o}$, needs to be smaller than the average value $\overline{SFR}$, i.e.  \begin{eqnarray}
    \frac{\overline{SFR}}{SFR_{\rm o}} ~>~ 1 \, .
    \label{eq:SFcond0}
\end{eqnarray} 
\citet{Dave08} suggests this to be broadly consistent with dark-matter-based hydrodynamical simulations. Galaxy-formation simulations in Milgromian gravitation based on single-cloud collapse with negligible further gas accretion also lead to such a general behaviour, albeit with a much larger $\overline{SFR}/SFR_{\rm o}$ \citep{Wittenburg+20}.

In order to improve knowledge of galaxy formation and evolution, it is useful to also consider the ensemble of galaxies in the nearby Universe. \citet{PeeblesNusser10} compare structure formation models to the spatial distribution of galaxies in the Local Cosmological Volume, defined by the sphere with radius $8\,$Mpc, finding significant tension between the predicted and observed distribution of major galaxies. Here, The Catalogue of Neighbouring Galaxies (Sec.~\ref{sec:LVG}) is used to address if $SFR_{\rm o} < \overline{SFR}$. In Sec.~\ref{sec:1:1}, it is shown that galaxies in the Local Cosmological Volume pose a challenge to this generally accepted understanding. Sec.~\ref{sec:conc} presents a discussion and the conclusions.

\section{The galaxies in the Local Cosmological Volume}
\label{sec:LVG}

The Catalogue of Neighbouring Galaxies \citep{Karachentsev+04} and its update \citep{Karachentsev+13} are used to extract the K-band luminosities and the SFRs based on integrated H$\alpha$ and far ultraviolet (FUV) measurements for galaxies within a distance of $\approx 11$~Mpc. The catalogue also lists limit flags on the FUV- and H$\alpha$-based SFR values. If a SFR value is marked with such a flag, it is excluded from the here presented analysis (but  including flagged SFR values does not significantly affect the results). This gives a sample of 870 galaxies, from 1145.
From these~870 galaxies, 267 have only far ultraviolet (FUV)-based SFR values, 128 have only H$\alpha$ based SFRs, and 475 have both measurements available. Thus, $14.7\%$ and $30.7\%$ galaxies lack FUV- and H$\alpha$-based SFR measurements, respectively.\footnote{The catalogue data are publicly available here:
\href{https://www.sao.ru/lv/lvgdb/tables.php}{https://www.sao.ru/lv/lvgdb/tables.php}. The
version updated on May 8th, 2020 is adopted here.}  The K-band luminosity values are converted to $M_*$ using a mass-to-light ratio of 0.6 \citep{Lelli+16}.

\begin{figure}
	\centering
	\includegraphics[width = 8.5cm, angle = 0]{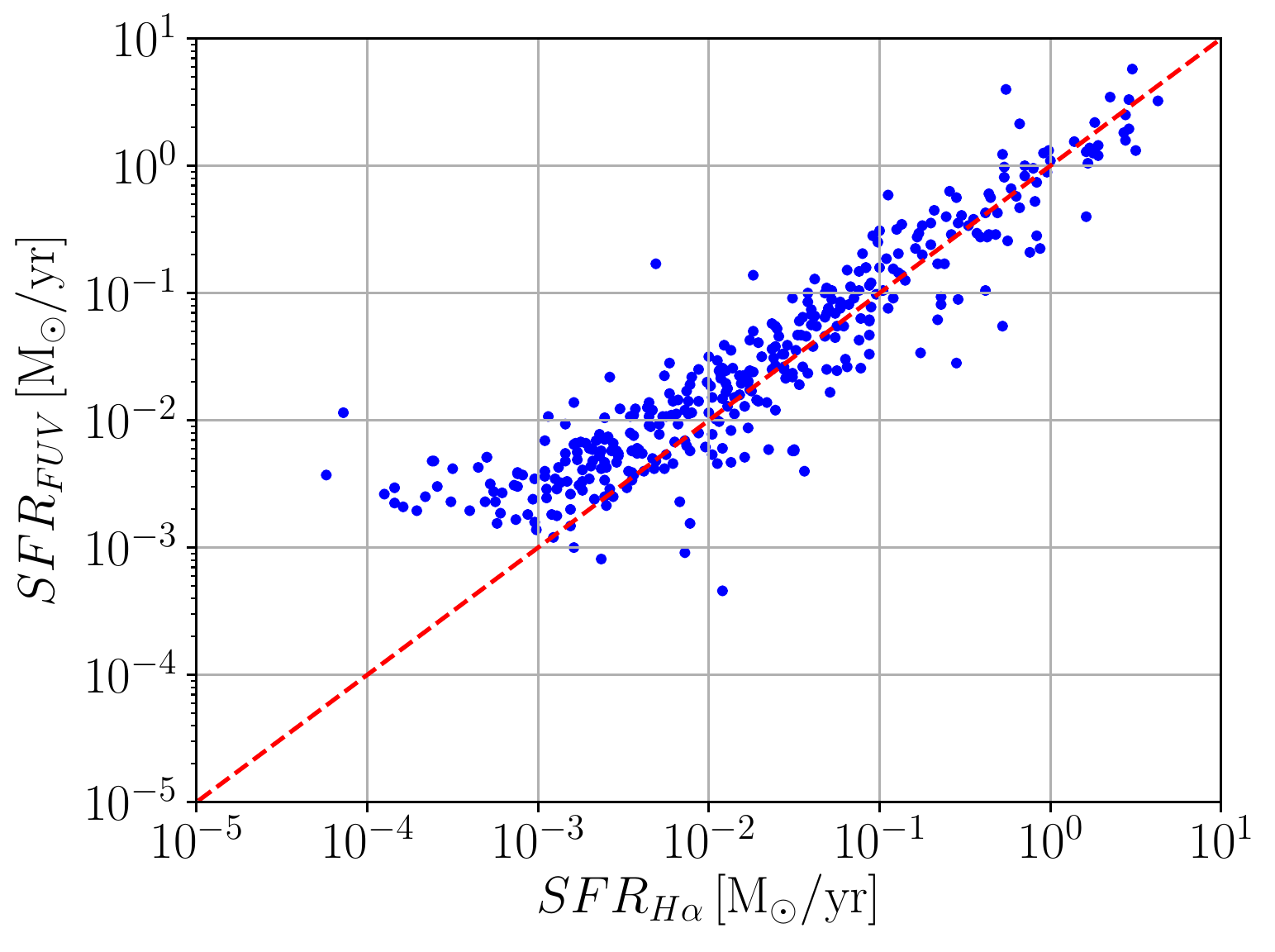}
	\caption{Comparison of the FUV- and H$\alpha$-based $SFR_{\rm o}$ values for those~386 galaxies in The Catalogue of Neighbouring Galaxies (Sec.~\ref{sec:LVG}) which have both measurements (and no limit flag on the SFR measurements) and for which also $SFR_{\rm o} = (SFR_{\rm FUV} + SFR_{\rm{H}\alpha})/2  \ge 10^{-3} \, M_\odot$/yr. The red dashed line is the 1:1 relation. The Catalogue of Neighbouring Galaxies does not provide uncertainties, but from the distance uncertainties \citep{Karachentsev+04, Karachentsev+13} it may be assumed that each measurement plotted here has a representative uncertainty of~30 per cent (not shown). Throughout the present analysis, the data are therefore treated with equal weights.}
	\label{fig:FUV_Halpha}
\end{figure}

For galaxies which have both SFR measurements available, $SFR_{\rm o} = (SFR_{\rm FUV} + SFR_{\rm{H}\alpha})/2$ is adopted. Their SFRs based on integrated H$\alpha$ and FUV measurements are depicted in Fig.~\ref{fig:FUV_Halpha}. The condition $SFR_{\rm o}\ge 10^{-3}\,M_\odot$/yr leaves~386 galaxies. For the galaxies which only have the FUV- or H$\alpha$-based values, $SFR_{\rm o} = SFR_{\rm FUV}$ or $SFR_{\rm o} = SFR_{{\rm H}\alpha}$, respectively (these galaxies do not appear in Fig.~\ref{fig:FUV_Halpha}). Applying the cut $SFR_{\rm o} \ge 10^{-3}\,M_\odot$/yr leaves~109 and~88 galaxies which have only $SFR_{\rm FUV}$ or $SFR_{{\rm H}\alpha}$ data, respectively. 

Galaxies with $SFR_{\rm o} < 10^{-3} \, M_\odot$/yr are not considered in the present analysis for the following four reasons (\citealt{KK13} provide an in-depth discussion of these issues):

1) The $SFR_{\rm o}$ values may be affected by a pronounced top-light galaxy-wide initial mass function (gwIMF) such that $SFR_{{\rm o, H}\alpha} < SFR_{\rm o, FUV}$.  \citet{PWK09,PK09} and sec.~4.3 in \citet{Jerabkova+18} provide a detailed discussion of this gwIMF behaviour, with the resolved dwarf galaxy DDO~154, which has a corresponding lack of massive stars, being a case in point \citep{Watts+18}. Indeed, the ratio $SFR_{{\rm o, H}\alpha}/SFR_{\rm o,FUV}$ is measured to decrease with decreasing $SFR_{\rm o, FUV}$ or $SFR_{{\rm o, H}\alpha}$ \citep{Lee+09}, as predicted via the IGIMF theory\footnote{The IGIMF theory is based on calculating the gwIMF by integrating over all star-forming events in a galaxy. These are localised in molecular cloud over-densities (cloud cores) containing at least a few binaries. The integrated galactic initial mass function (IGIMF) is thus the galaxy-wide integral over all freshly formed embedded clusters. The reader is referred to \cite{Kroupa+13} and \cite{Hopkins18} for a discussion of this problem.} if the gwIMF becomes increasingly top-light with decreasing SFR \citep{PWK09}, a trend observationally also suggested for more massive disk galaxies \citep{Gunawardhana+11}. This behaviour of the gwIMF with $SFR$ would imply dwarf galaxies to be as efficient in forming stars as major disk galaxies (all having a gas depletion or consumption time-scale of about 3~Gyr), therewith explaining their ability to form their observed stellar populations within a Hubble time without the need to postulate that they have had much higher SFRs in the past \citep{PWK09}. Also noteworthy in this context is that two well studied galaxies, the ultra-diffuse Dragonfly~44 and the ultra-faint satellite Bo\"otes~I, had the same gas-depletion time as the above-mentioned star-forming dwarf galaxies as long as the analysis is performed self-consistently within the IGIMF theory \citep{Haghi+19, Yan+20}.

2) Related to the systematically changing gwIMF just discussed, it has been customary to describe star formation throughout a galaxy as being stochastic, an approach discussed in detail in \cite{Elmegreen97, Elmegreen99}.\footnote{The present authors are not suggesting that stars can appear in a vacuum although the notion of stochastic star formation needs to be understood in this context, see \cite{Elmegreen97, Elmegreen99} for further details.} Dwarf galaxies with low SFRs may lead to an undersampling of the gwIMF, but only if a central tenet of the IGIMF theory is adopted, namely that stars form in embedded clusters which dictate the available mass for the stellar population per cluster, leading to a difference in various SFR tracers \citep{daSilva+14}. Related to the suggestion that star formation is stochastic, \cite{KE12} discuss the possibility that dwarf galaxies may be prone to temporal variations of their SFRs. \cite{Applebaum+20} study the effect of a stochastic gwIMF on the 
formation and evolution of dwarf galaxies with cosmological simulations.

3) It is possible that the measured H$\alpha$ and UV fluxes are affected by photon leakage, dust obscuration and redenning and other physical effects \citep{Calzetti08, Calzetti13}.  It is likely that these are relevant, but the authors of the research papers reporting evidence for systematic variations of the gwIMF in galaxies correct for these biases \citep{HG08,Lee+09, Gunawardhana+11}.

4) Many of the least-massive galaxies in the Local Cosmological Volume with $M_* < 10^7 \, M_\odot$ are satellite galaxies which stopped forming stars many~Gyr ago \citep[e.g.][]{Grebel97, Grebel+03}. Dwarf galaxies sufficiently far from major galaxies of this mass behave as disk galaxies \citep[e.g. the dwarf galaxy DDO~210 which is barely more massive than a massive globular cluster and which has a Milgromian rotation curve, putting it on the baryonic-Tully-Fisher relation,][]{FamaeyMcGaugh12}. Meanwhile, dwarf galaxies within the vicinity of major galaxies have lost their gas \citep{Grebel+03}. 

Overall, dwarf and major star-forming galaxies thus appear to conform to the same basic physical evolution, although differences between the SFHs at the lowest-mass end are evident and remain to be understood \citep{Albers+19}.  Galaxies with $SFR_{\rm o}< 10^{-3}\,M_\odot$/yr are therefore considered to have too insecure SFRs and are thus omitted from the present analysis.  For the in-total~583 galaxies used in this analysis, the gwIMF is near to canonical \citep{Gunawardhana+11, Jerabkova+18}.

In a companion paper (Haslbauer et al., in prep.),  The Catalogue of Neighbouring Galaxies is compared  with the \citetalias{Speagle+14} main sequence. The $\left( SFR_{\rm o}, M_* \right)$ values are in excellent agreement at $M_*>10^9\,M_\odot$, showing that The Catalogue of Neighbouring Galaxies conforms to the compilation by \citetalias{Speagle+14}.

\section{The current vs the average SFR}
\label{sec:1:1}

\subsection{Theoretical expectations}
\label{Theoretical_expectations}

Assume a galaxy forms stars from time $t_1$ until time $t_2$, the latter being the present-day if the galaxy is still forming stars. The SFH can be written as $SFR \left( t \right) = \overline{SFR} + \Delta SFR \left( t \right)$ such that the average SFR of any galaxy over the time scale $t_{\rm sf} \equiv t_2 - t_1$ becomes
\begin{eqnarray}
    \overline{SFR} ~=~ \frac{1}{t_{\rm sf}}\, \left( \int_{t_1}^{t_2} \overline{SFR} \, dt + \int_{t_1}^{t_2} \Delta SFR \left( t \right) \, dt \right) \, ,
    \label{eq:avSFR}
\end{eqnarray}
with temporal deviations from $\overline{SFR}$ satisfying,
\begin{eqnarray}
    \int_{t_1}^{t_2} \Delta SFR \left( t \right) \, dt ~=~ 0 \, M_\odot \, .
    \label{eq:dSFR}
\end{eqnarray}
This implies that if $SFR_{\rm o} = \overline{SFR}$, any decreasing SFH must have had a compensating increase.

\begin{figure}
	\centering
	\includegraphics[width = 8.5cm, angle = 0]{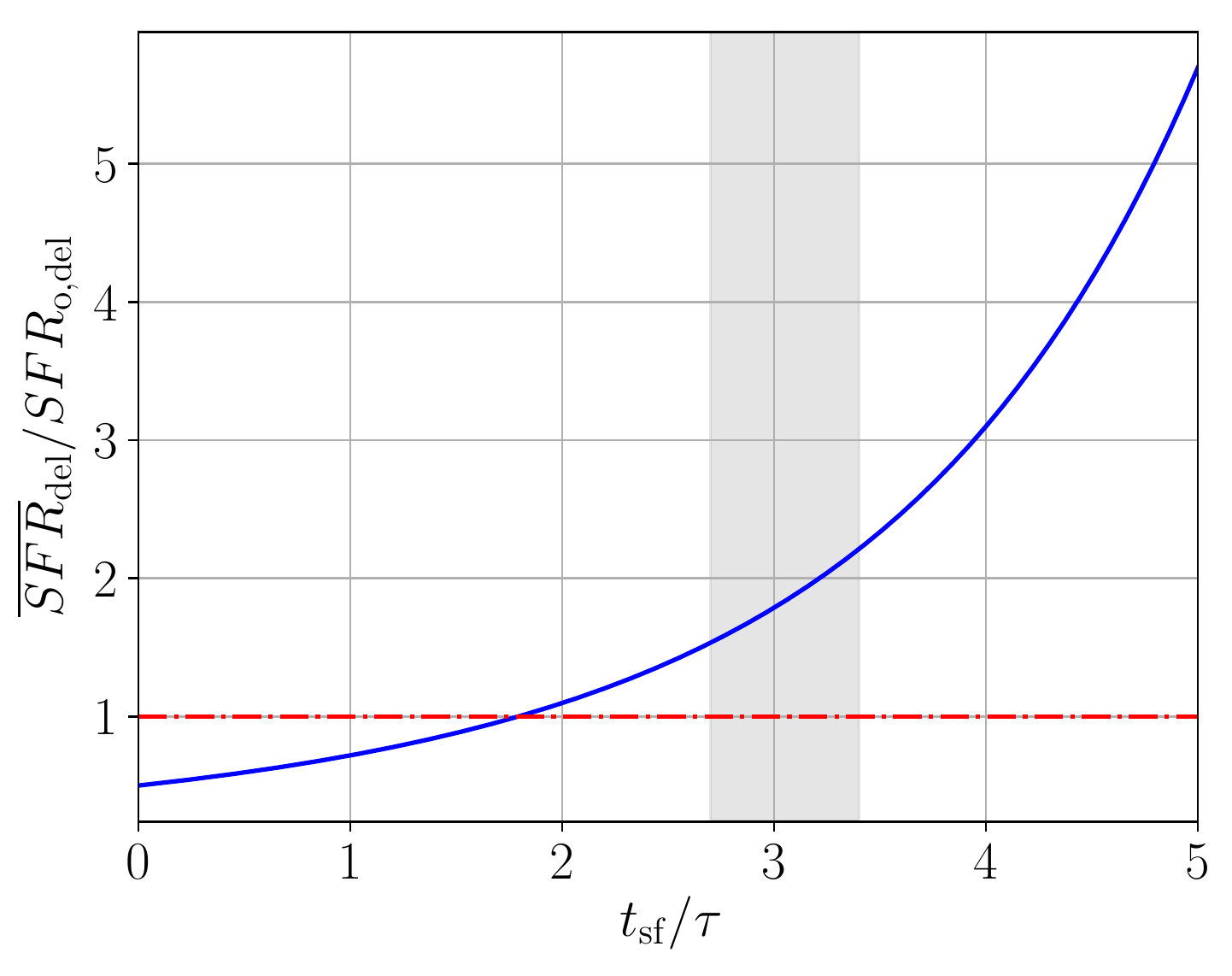}
	\caption{The ratio between the average and present-day SFR, $\overline{SFR}/SFR_{\rm o}$ (Eq.~\ref{eq:SFcond}) as a function of $t_{\rm sf}/\tau$. The shaded vertical band contains galaxies with $2.7 < t_{\rm sf}/\tau<3.4$ (Eq.~\ref{eq:SFcond_Speagle}).}
	\label{fig:rat}
\end{figure}

According to the analysis by \citetalias{Speagle+14}, the SFHs of galaxies can be described by ``delayed-$\tau$'' models of the form
\begin{eqnarray}
    SFR_{\rm del}\left( t > t_1 \right) ~=~ \frac{A_{\rm del}}{\tau^2} \, (t-t_1) \, \exp \left(-\frac{t - t_1}{\tau} \right) \, ,
    \label{eq:deltau}
\end{eqnarray}
where $A_{\rm del}$ is a normalisation constant and $\tau$ is the star-formation time scale, with $SFR(t \le t_1)=0\,M_\odot$/yr. Note that the present-day SFR becomes
\begin{eqnarray}
    SFR_{\rm del}(t_2) &\equiv& SFR_{\rm o, del} ~=~ \frac{A_{\rm del} \, x \, \mathrm{e}^{-x}}{\tau} ~\text{, where} \\
  x &\equiv& \frac{t_{\rm sf}}{\tau} \, .
\end{eqnarray}
The parameter $x$ signifies which phase of the SFH a galaxy is currently in $-$ the SFH rises up to a peak at $x = 1$ before following an asymptotically exponential decline. With this SFH, the average SFR is
\begin{eqnarray}
    \overline{SFR}_{\rm del} ~=~ \frac{A_{\rm del}}{t_{\rm sf}} \, \left[1 - \left( 1 + x \right) \, \mathrm{e}^{-x} \right] \, .
\end{eqnarray}
Thus \citepalias[see also fig.~10 in][]{Speagle+14},
\begin{eqnarray}
	\frac{\overline{SFR}_{\rm del}}{SFR_{\rm o,del}} ~=~ \frac{\mathrm{e}^x - x - 1}{x^2} ~\geq~ \frac{1}{2} \, ,
	\label{eq:SFcond}
\end{eqnarray}
the value of $\frac{1}{2}$ being reached at $x = 0$ (very long $\tau$). The ratio (Eq.~\ref{eq:SFcond}) is shown in Fig.~\ref{fig:rat} as a function of $x$, revealing that the regime in which $\overline{SFR}/SFR_{\rm o}>1$ occurs for $x > 1.79$.

The observationally constrained galaxy main-sequence evolution analysed by \citetalias{Speagle+14} suggests that the typical real star-forming galaxy undergoes an evolution beginning at $t_1\approx 3\,$Gyr and with $SFR_{\rm del,max}/SFR_{\rm o,del} > 2$. The maximum value of the SFR, $SFR_{\rm del, max}$, is reached when $t-t_1=\tau$ at the time $6.5 < t/{\rm Gyr} < 7.5$ \citepalias[fig.~10 in][]{Speagle+14} such that $3.5< \tau/{\rm Gyr} < 4.5$. Since $t_{\rm sf}\approx 12\,$Gyr for the galaxies observed in the Local Group, these galaxies would be in the $2.7 < x < 3.4$ phase of their evolution, i.e. in the falling part of their SFH.  If the nearby galaxies concur to this evolution, then according to Fig.~\ref{fig:rat} they ought to have
\begin{eqnarray}
  1.5 < \frac{\overline{SFR}}{SFR_{\rm o}} < 2.3 \, .
  \label{eq:SFcond_Speagle}
\end{eqnarray} 
Note that $2.1 < \overline{SFR}/SFR_{\rm o} < 2.6$ if $t_1=1.8\,$Gyr (i.e. if galaxies started to form stars 12$\,$Gyr ago), worsening the problem.

\begin{figure}
	\centering
	\includegraphics[width = 8.5cm, angle = 0]{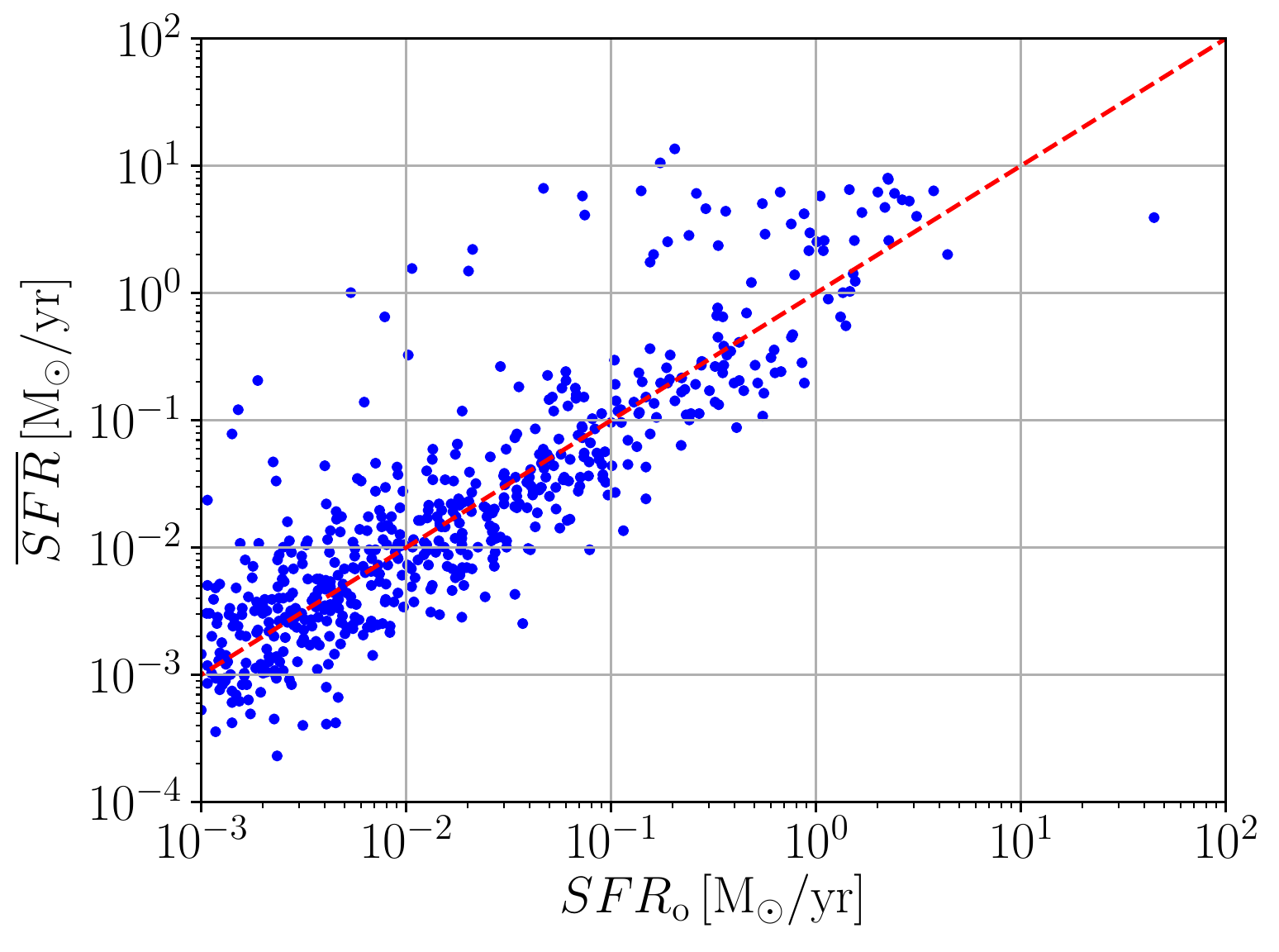}
	\caption{The present-day measured SFR, $SFR_{\rm o}$, is compared with the average value, $\overline{SFR}$, for each galaxy (filled squares, Sec.~\ref{sec:LVG}) for galaxies with $SFR_{\rm o} \ge 10^{-3}\,M_\odot$/yr and assuming $\zeta = 1.3, t_{\rm sf} = 12\,$Gyr. The dashed line is the 1:1 relation. Note that the data suggest a second sequence of galaxies lying $\approx 2$~dex above the ridge line populated by the majority of galaxies near the 1:1 line. This sequence is discussed further in Haslbauer et al. (in prep.) and may be related to galaxies which have lost their inter-stellar medium being replenished through stellar evolutionary mass loss.}
	\label{fig:1:1}
\end{figure}

\subsection{The Local Volume}
\label{Local_volume}

The sample of Neighbouring Galaxies (Sec.~\ref{sec:LVG}) can be used to assess the ratio $\overline{SFR}/SFR_{\rm o}$ because these galaxies have well-measured $M_*$ and $SFR_{\rm o}$ values. The present-day stellar mass of the galaxy defines the average SFR,
\begin{eqnarray}
    \overline{SFR} ~\equiv~ \frac{\zeta \, M_*}{t_{\rm sf}} \, ,
    \label{eq:Mstar}
\end{eqnarray}
where $\zeta$ accommodates mass loss through stellar evolution \citep[$\zeta \approx 1.3$ according to a canonical single-burst stellar population, as evident in fig.~1 of][]{BaumgardtMakino03}. Note that $1 < \zeta < 1.3$ for a galaxy with a constant SFR and a canonical gwIMF, but $\zeta$ also depends on the gwIMF and can reach values as large as 2--3 for extreme star-bursting, low-metallicity galaxies \citep[according to the IGIMF theory, see e.g.][]{Yan+17, Dabring19, Yan+19}. According to the IGIMF theory, the nearby galaxies used here are expected to have $1.0 < \zeta < 1.3$. By adopting $\zeta=1.3$ we are thus working conservatively $-$ the expected smaller value would bring a larger conflict with the delayed-$\tau$ SFHs (see below).

\begin{figure}
	\centering
	\includegraphics[width = 8.5cm, angle = 0]{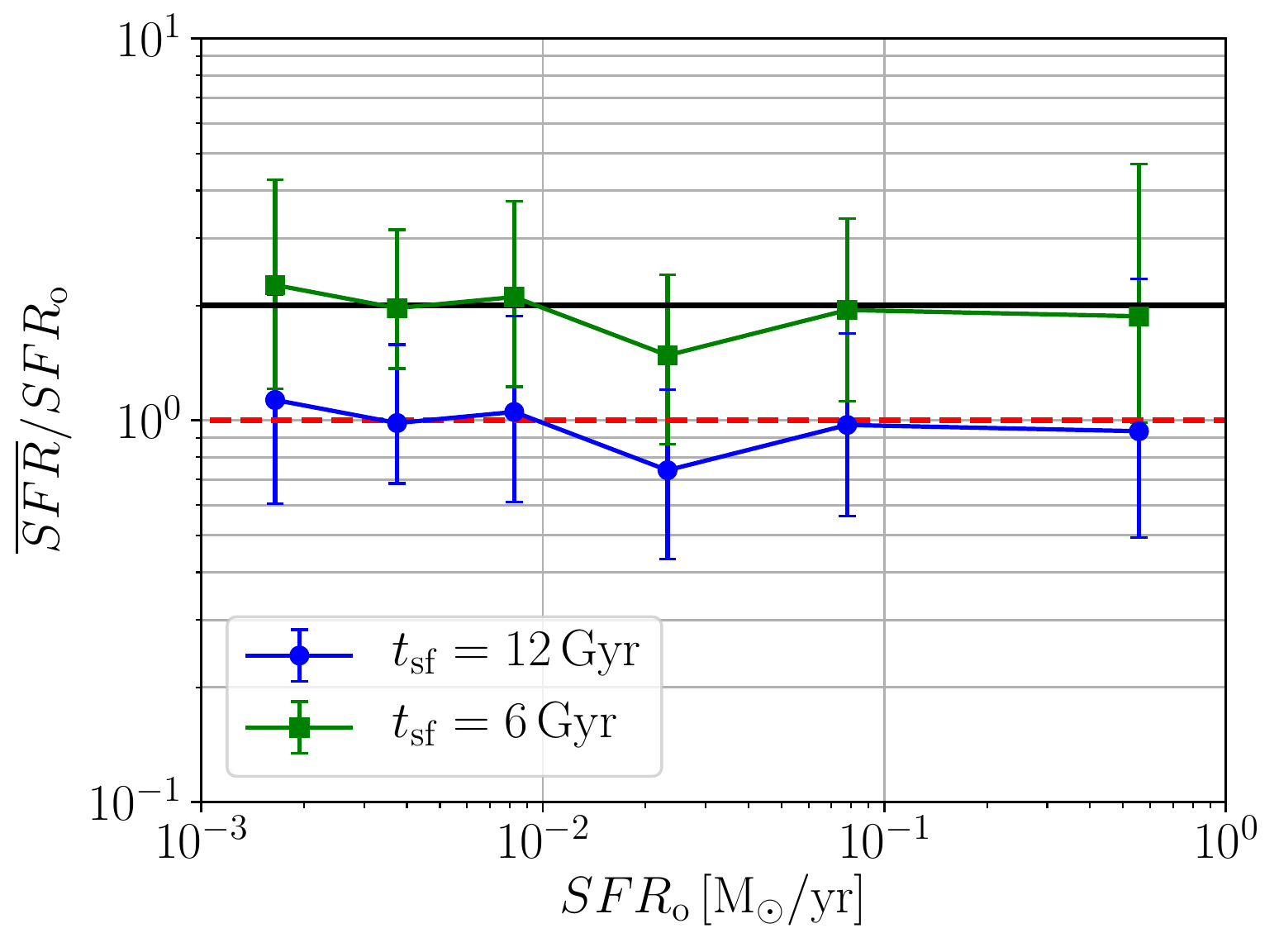}
	\caption{Using galaxies with $SFR_{\rm o} \ge 10^{-3}\,M_\odot$/yr, the ratio between the average SFR and the present-day value is plotted in bins containing~100 galaxies (except for the right-most bin, which contains~83 galaxies) in order to test Eq.~\ref{eq:SFcond0}. The points and error bars show the median and first-third interquartile range, respectively. The blue (green) points (squares) are for $t_{\rm sf} = 12\,$(6)$\,$Gyr, with lower $t_{\rm sf}$ leading to higher $\overline{SFR}$. The solid black line depicts the expected value of 2 (Eq.~\ref{eq:SFcond_Speagle}), while the dashed red line shows a value of~1. } 
\label{fig:dat_rat}
\end{figure}

Both observationally derived values, $SFR_{\rm o}$ and $\overline{SFR}$, are compared in  Fig.~\ref{fig:1:1} for the galaxies with $SFR_{\rm o} \ge  10^{-3}\,M_\odot$/yr. Fig.~\ref{fig:dat_rat} plots the empirical $\overline{SFR}/SFR_{\rm o}$ ratios to test if Eq.~\ref{eq:SFcond0} or~\ref{eq:SFcond_Speagle} can be fulfilled by the data. It is apparent that the delayed-$\tau$ SFH description cannot account for the data if $t_{\rm sf}=12\,$Gyr and $\tau \approx 4$~Gyr. Reducing $t_{\rm sf}$ to 6 Gyr would allow the model to account for the data, but such short $t_{\rm sf}$ violates the observed ages of galactic disks, which suggest $t_{\rm sf}\approx 12\,$Gyr \citep[e.g.][]{Knox_1999}.

\begin{figure}
	\centering
	\includegraphics[width = 8.5cm, angle = 0]{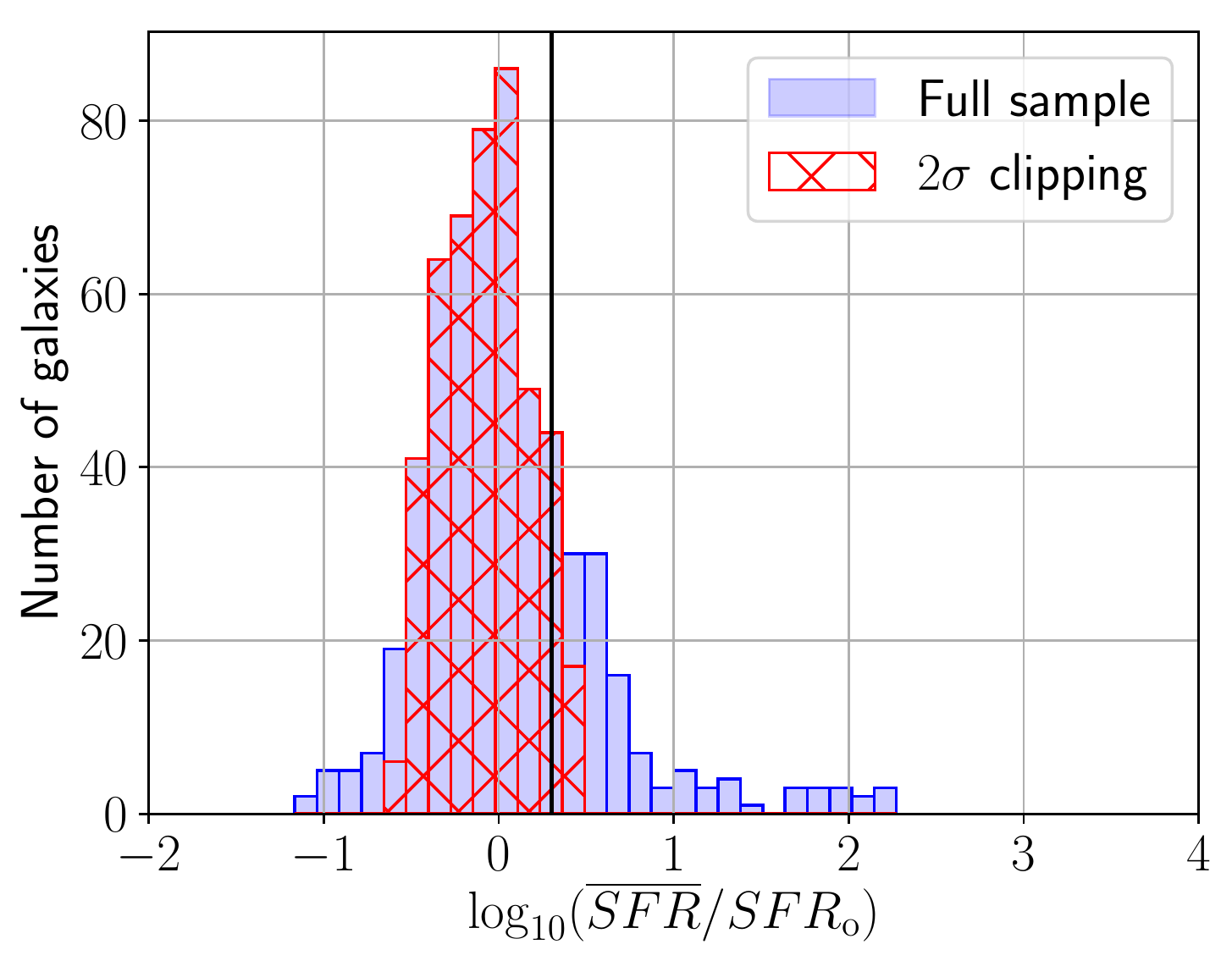}
 \caption{Histogram of $\overline{SFR}/SFR_{\rm o}$ values for galaxies with $SFR_{\rm o} \ge 10^{-3} M_\odot/{\rm yr}$ if $t_{\rm sf}=12\,$Gyr. The hatched red region shows the effect of $2\sigma$ outlier rejection, which reduces the sample size from~583 to~455. The $\overline{SFR}/SFR_{\rm o}$ values would double if $t_{\rm sf}=6\,$Gyr, so for clarity the expected horizontal shift is plotted as a vertical solid black line at $\log_{10} 2$. This is also the expected value according to the SFH obtained by \citetalias{Speagle+14} (Eq.~\ref{eq:SFcond_Speagle}).}
	\label{fig:hist}
\end{figure}

As is evident from Fig.~\ref{fig:1:1}, in the Local Cosmological Volume, $\overline{SFR}/SFR_{\rm o}$ is pushed to larger values because of the second upper sequence (evident at $\log_{10}\left( \overline{SFR}/SFR_{\rm o} \right) = 2$ in Fig.~\ref{fig:hist}). By removing the data in the second sequence using $2\sigma$ clipping, an improved estimate of the ratio can be obtained for the majority of nearby star-forming galaxies. The dispersion of the ratios for the galaxies with $SFR_{\rm o} \ge 10^{-3}\,M_\odot/$yr is calculated and all data which are further than $2\sigma$ from the average are taken out. This is repeated until the average ratio does not shift, removing the second sequence noted in Fig.~\ref{fig:1:1}. The clipped and non-clipped distributions are displayed in Fig.~\ref{fig:hist}.

Prior to outlier rejection, the sample of~583 galaxies yields a median (mean) $\left( \overline{SFR}/SFR_{\rm o} \right)$ of 0.96 (1.11).\footnote{The median and mean values and outlier rejection are calculated in log$_{10}$ space.} Applying outlier rejection leaves~455 galaxies and yields a median (mean) $\left( \overline{SFR}/SFR_{\rm o} \right)_{\rm clip}$ of 0.84 (0.85). 
The dispersion is 0.243~dex, but given the number of galaxies, the uncertainty in the mean is formally only~0.011 dex. 
The uncertainty of the mean of $\left( \overline{SFR}/SFR_{\rm o} \right)$ is calculated by
\begin{eqnarray}
\sigma = \frac{0.85}{2} \bigg( 10^{0.011} - 10^{-0.011} \bigg) \approx 0.02 \, .
\end{eqnarray}
For completeness, it is noted that including also the~586 galaxies with $SFR_{\rm o}\ge 10^{-3}\,M_\odot$/yr
which are marked in The Catalogue of Neighbouring Galaxies with a limit flag ($2\,\sigma$ clipping leaves 472 galaxies from this sample), the median (mean) of $\left( \overline{SFR}/SFR_{\rm o} \right)_{\rm clip}$ becomes $0.84 \pm 0.02$ ($0.85 \pm 0.02$), which is not distinguishable from the main result.

There is a clear preference for $\left( \overline{SFR}/SFR_{\rm o} \right)_{\rm clip} < 1$, though given the possibility of systematics (e.g. in the mass-to-light ratio), $\left( \overline{SFR}/SFR_{\rm o} \right)_{\rm clip} = 1$ is possible, in which case $x=1.79$ and $\tau=6.7\,$Gyr assuming $t_{\rm sf}=12\,$Gyr. But the data certainly argue against $\left( \overline{SFR}/SFR_{\rm o} \right)_{\rm clip} > 1$.

\subsection{A power-law SFH?}
\label{Power_law_SFH}

Which type of SFH might account for the observational data and in particular for the mean value of $\left( \overline{SFR}/SFR_{\rm o} \right)_{\rm clip} = 0.85$, or at least get the ratio down to 1? As an ansatz, 
\begin{eqnarray}
    SFR_{\rm incr} \left( t > t_1 \right) ~=~ A_{\rm incr} \, \left( t - t_1 \right)^\eta \, , 
    \label{eq:incrSF}
\end{eqnarray}
with $SFR_{\rm incr} \left( t \le t_1 \right) = 0 \, M_\odot$/yr as before, $\eta \ne -1$ being the power-law index and $A_{\rm incr}$ the normalisation.

The present-day average and present-day SFR are, respectively,
\begin{eqnarray}
	\overline{SFR}_{\rm incr} &=& \frac{A_{\rm incr}}{1 + \eta}\,t_{\rm sf}^\eta \, , \\
    SFR_{\rm o,incr} &=& A_{\rm incr}\,t_{\rm sf}^\eta \, .
\end{eqnarray}
To get the correct total normalisation $M_*$ at the present time, 
\begin{eqnarray}
	A_{\rm incr} ~=~ \zeta\,\left( 1 + \eta \right) \, M_*/ t_{\rm sf}^{1 + \eta} \, .
\end{eqnarray}
Thus,
\begin{eqnarray}
    \frac{\overline{SFR}_{\rm incr}}{SFR_{\rm o,incr}} ~=~ \frac{1}{1 + \eta} \le 1 \, .
\end{eqnarray}
The uncertainty of $\eta$ is
\begin{eqnarray}
\sigma_{\eta} ~&=&~ \frac{\eta_{-} - \eta_{+}}{2} \, , \qquad \mathrm{where} \\
\eta_{\pm} ~&=&~ \frac{1}{0.85 \times 10^{\pm0.011}} - 1 \, .
\end{eqnarray}
The mean value of $\left( \overline{SFR}/SFR_{\rm o} \right)_{\rm clip} = 0.85 \pm 0.02$ (Section \ref{Local_volume}), implying that $\eta = 0.18 \pm 0.03$. The true uncertainty is probably at least about~30~per cent larger since the value quoted here treats all data points with equal weight and ignores the observational uncertainties which are not provided by \cite{Karachentsev+04, Karachentsev+13}.  Note that $\overline{SFR}_{\rm incr}/SFR_{\rm o,incr} = 1$ can be recovered for the case $\eta = 0$, which implies that any individual galaxy has the same SFR at all times. 

According to this model, the stellar mass growth of an isolated non-interacting galaxy would be
\begin{eqnarray}
    M_*\left( t - t_1 \right) ~=~ \frac{1}{\zeta \left( 1 + \eta \right)} \, \left(t - t_1 \right) \, SFR \left( t - t_1 \right) \, .
    \label{eq:incrMgrowth}
\end{eqnarray}
For the case $\eta = 0$, $M_* \propto \left( t - t_1 \right)$. Since the cosmic scale factor $a \left( t \right) \appropto t$, this model would be very similar to the case $M_* \propto a$ at late times, where the observational constraints are strongest.

Assuming all galaxies begin forming stars at the same $t_1$, the main sequence would thus evolve according to 
\begin{eqnarray}
    SFR \left( t - t_1 \right) ~=~ \zeta \, \left(1+\eta\right)\, \frac{M_*\left( t - t_1 \right)}{t - t_1} \, .
    \label{eq:incr_MS}
\end{eqnarray}
At a given $M_*$, the SFR would decrease with an increasing age of the Universe, $t$, because a galaxy with fixed $M_*$ needed a larger SFR to reach this mass in a shorter time when the Universe was younger.

As a quantitative example, consider galaxies with a fixed $M_*=10^{10}\,M_\odot$ at different times in standard cosmology. Suppose these all started forming stars at $t_1 = 1.8$~Gyr (i.e. about $12\,$Gyr ago) according to $\zeta = 1.3, \eta = 0$. At $t = 2.1$~Gyr, $z \approx 3$ such that $SFR \left( 0.3\,{\rm Gyr} \right) = 43 \, M_\odot$/yr. At $t = 6\,$Gyr, $z\approx1$ and $SFR = 3.1\,M_\odot$/yr. These are broadly consistent with the evolution documented by \citetalias{Speagle+14}, namely that $SFR\approx 32^{+24}_{-18}\,M_\odot$/yr at $z = 3$ and $SFR\approx 10^{+7.8}_{-5.6}\,M_\odot$/yr at $z = 1$ (their fig.~8).

The specific SFR becomes
\begin{eqnarray}
    sSFR \left( t - t_1 \right) ~=~ \zeta \, \left( 1 + \eta \right) \frac{1}{t - t_1} \, ,
    \label{eq:incr_sSFR}
\end{eqnarray}
such that today ($t - t_1 = t_{\rm sf} = 12\,$Gyr and with $\zeta=1.3, \eta=0$),
\begin{eqnarray}
    sSFR \left( t - t_1 \right) ~=~ 1.1\times 10^{-10} \, {\rm yr}^{-1} \, .
    \label{eq:incr_sSFR_today}
\end{eqnarray}
As a further consistency check, in the regime where the gwIMF is expected to be comparable to the canonical IMF (i.e. $\zeta \approx 1.3$) and thus for $M_*=10^{10}\,M_\odot$, the observational study by \cite{Ilbert+15} leads to $sSFR=1.23\times 10^{-10}\,$yr$^{-1}$ (their eq.~1 with $z=0$), being in reasonable agreement with~Eq.~\ref{eq:incr_sSFR_today}. 
Note that the $sSFR$ vs $M_*$ relation might not be constant as its slope would be given by the value $\zeta\left(SFR_{\rm o}, Z\right)$ if the gwIMF depends on the SFR and metallicity, $Z$. According to the IGIMF theory, $\zeta > 1.3$ for galaxies with $SFR_{\rm o}>10\,M_\odot$/yr \citep[cf.][]{Gunawardhana+11,Yan+17}.  The slope of the $sSFR$ vs $M_*$ relation thus encodes the systematic variation of the gwIMF.

\section{Discussion and Conclusion}
\label{sec:conc}

The Catalogue of Neighbouring Galaxies (Sec.~\ref{sec:LVG}) provides a sample of star forming galaxies which are used here to test a robust expectation concerning galactic SFHs (Eqs.~\ref{eq:SFcond0} and \ref{eq:SFcond_Speagle}). This expectation comes from the general understanding that galaxies typically have decreasing SFRs now, and should thus have an average star formation rate exceeding the present value ($\overline{SFR}/SFR_{\rm o} > 1$). Within the context of the delayed-$\tau$ models (Section \ref{Theoretical_expectations}), the present results are indicative of a rather long star formation timescale of $\tau \approx 6.7$~Gyr, much longer than the $3.5-4.5$~Gyr estimated by \citetalias{Speagle+14}. The tension between their results and the local sample is also evident in Figs.~\ref{fig:1:1} and \ref{fig:hist}, which show that for the plausible assumption that galaxies have been forming stars for 12 Gyr, the average and present SFRs are nearly the same. However, $\overline{SFR}/SFR_{\rm o} \approx 2$ is expected (Fig.~\ref{fig:rat}).

The nearby galaxies with $SFR_{\rm o} \ge 10^{-3}\,M_\odot$/yr are in tension with this expectation by having present-day SFRs which are larger than the average values as estimated from the observationally constrained masses in stars. The data are consistent with galaxies having a constant SFH ($\eta = 0$ in Eq.~\ref{eq:incrSF}). While the mass of an individual galaxy increases with time, for a given fixed stellar mass, the SFR would decrease with time (Eq.~\ref{eq:incr_MS}). This is broadly consistent with the results obtained by \citetalias{Speagle+14}. The slope of the $sSFR$ vs $M_*$ relation reflects the systematically varying gwIMF.

A caveat of these deductions is that the data in The Catalogue of Neighbouring Galaxies may be systematically biased, for example if the measured SFRs are too large or the observationally derived stellar masses are too low. However, the bias would need to be a factor of $\approx 1.2$ in $SFR_{\rm o}$ or in $M_*$ (in this case the K-band mass-to-light ratio would be~0.72 rather than~0.6 as used in Sec.~\ref{sec:LVG}) to shift $\overline{SFR}/SFR_{\rm o}$ to the regime $>1$ (Sec.~\ref{sec:1:1}). While this could occur, a much larger bias would be required to get the expected ratio of $\overline{SFR}/SFR_{\rm o} \approx 2$ (Sec.~\ref{Theoretical_expectations}). This appears implausible given that these galaxies comprise the nearby sample, which ought to be well observed. Furthermore, the $SFR_{\rm o}$ and $M_*$ values are consistent with the redshift $z = 0$ main sequence as determined by \citetalias{Speagle+14} and shown explicitly in Haslbauer et al. (in preparation). 

It may also be possible that the Local Cosmological Volume, by being part of a 400~Mpc underdensity \citep{Karachentsev12, Keenan+13}, is not entirely representative and that the galaxies it contains lag in their evolution compared to galaxies further away, which are used by \citetalias{Speagle+14} to quantify the evolution of the main sequence. The star-by-star data of the Local Group galaxies and their globular clusters \citep{Grebel97} do not, however, support the notion that the Local Cosmological Volume contains younger galaxies than the cosmological average.

If the tension between the Neighbouring Galaxies and the evolution of the galactic SFHs \citepalias[fig.~10 in][]{Speagle+14} remains, then this may be indicative of an issue with calculating the rest-frame physical properties of galaxies at $z > 0$ and thus with the cosmological model employed for this calculation \citep[cf.][]{BK19}. It may be worth exploring which cosmological model might ease the tension, but this is beyond the scope of the present contribution.

Assuming that the data in the Local Cosmological Volume used here are robust and unbiased, then they would possibly suggest a roughly constant or slightly increasing SFR in a typical galaxy over time. The flocculent disk galaxy Triangulum indeed suggests an increasing SFR. The physical origin of this dependency of the SFR on time and present-day mass of the galaxy (Eq.~\ref{eq:incrSF}) remain to be explored in this eventuality.

\section*{Acknowledgements}

IB is supported by an Alexander von Humboldt postdoctoral research fellowship. We thank Tereza Jerabkova for useful discussions.
PK acknowledges support from the Grant Agency of the Czech Republic
under grant number 20-21855S. This work benefited from the
International Space Science Institute (ISSI/ISSI-BJ) in Bern and
Beijing, thanks to the funding of the team “Chemical abundances in the
ISM: the litmus test of stellar IMF variations in galaxies across
cosmic time” (Donatella Romano and Zhi-Yu Zhang).

\section*{Data Availability Statement}
The data used stem from the Catalogue of Neighbouring Galaxies (Sec.~\ref{sec:LVG}). 

\bibliographystyle{mnras}
\bibliography{references}

\begin{thebibliography}{}
\makeatletter
\relax
\def\mn@urlcharsother{\let\do\@makeother \do\$\do\&\do\#\do\^\do\_\do\%\do\~}
\def\mn@doi{\begingroup\mn@urlcharsother \@ifnextchar [ {\mn@doi@}
  {\mn@doi@[]}}
\def\mn@doi@[#1]#2{\def\@tempa{#1}\ifx\@tempa\@empty \href
  {http://dx.doi.org/#2} {doi:#2}\else \href {http://dx.doi.org/#2} {#1}\fi
  \endgroup}
\def\mn@eprint#1#2{\mn@eprint@#1:#2::\@nil}
\def\mn@eprint@arXiv#1{\href {http://arxiv.org/abs/#1} {{\tt arXiv:#1}}}
\def\mn@eprint@dblp#1{\href {http://dblp.uni-trier.de/rec/bibtex/#1.xml}
  {dblp:#1}}
\def\mn@eprint@#1:#2:#3:#4\@nil{\def\@tempa {#1}\def\@tempb {#2}\def\@tempc
  {#3}\ifx \@tempc \@empty \let \@tempc \@tempb \let \@tempb \@tempa \fi \ifx
  \@tempb \@empty \def\@tempb {arXiv}\fi \@ifundefined
  {mn@eprint@\@tempb}{\@tempb:\@tempc}{\expandafter \expandafter \csname
  mn@eprint@\@tempb\endcsname \expandafter{\@tempc}}}

\bibitem[\protect\citeauthoryear{{Albers} et~al.,}{{Albers}
  et~al.}{2019}]{Albers+19}
{Albers} S.~M.,  et~al., 2019, \mn@doi [\mnras] {10.1093/mnras/stz2903}, \href
  {https://ui.adsabs.harvard.edu/abs/2019MNRAS.490.5538A} {490, 5538}

\bibitem[\protect\citeauthoryear{{Applebaum}, {Brooks}, {Quinn}  \&
  {Christensen}}{{Applebaum} et~al.}{2020}]{Applebaum+20}
{Applebaum} E.,  {Brooks} A.~M.,  {Quinn} T.~R.,   {Christensen} C.~R.,  2020,
  \mn@doi [\mnras] {10.1093/mnras/stz3331}, \href
  {https://ui.adsabs.harvard.edu/abs/2020MNRAS.492....8A} {492, 8}

\bibitem[\protect\citeauthoryear{{Balakrishna Subramani}, {Kroupa}, {Shenavar}
  \& {Muralidhara}}{{Balakrishna Subramani} et~al.}{2019}]{BK19}
{Balakrishna Subramani} V.,  {Kroupa} P.,  {Shenavar} H.,   {Muralidhara} V.,
  2019, \mn@doi [\mnras] {10.1093/mnras/stz2027}, \href
  {https://ui.adsabs.harvard.edu/abs/2019MNRAS.488.3876B} {488, 3876}

\bibitem[\protect\citeauthoryear{{Banik}, {O'Ryan}  \& {Zhao}}{{Banik}
  et~al.}{2018}]{Banik+18}
{Banik} I.,  {O'Ryan} D.,   {Zhao} H.,  2018, \mn@doi [\mnras]
  {10.1093/mnras/sty919}, \href
  {https://ui.adsabs.harvard.edu/abs/2018MNRAS.477.4768B} {477, 4768}

\bibitem[\protect\citeauthoryear{{Baumgardt} \& {Makino}}{{Baumgardt} \&
  {Makino}}{2003}]{BaumgardtMakino03}
{Baumgardt} H.,  {Makino} J.,  2003, \mn@doi [\mnras]
  {10.1046/j.1365-8711.2003.06286.x}, \href
  {https://ui.adsabs.harvard.edu/abs/2003MNRAS.340..227B} {340, 227}

\bibitem[\protect\citeauthoryear{{Bekenstein} \& {Milgrom}}{{Bekenstein} \&
  {Milgrom}}{1984}]{BM84}
{Bekenstein} J.,  {Milgrom} M.,  1984, \mn@doi [\apj] {10.1086/162570}, \href
  {https://ui.adsabs.harvard.edu/abs/1984ApJ...286....7B} {286, 7}

\bibitem[\protect\citeauthoryear{{B{\'\i}lek}, {Thies}, {Kroupa}  \&
  {Famaey}}{{B{\'\i}lek} et~al.}{2018}]{Bilek+18}
{B{\'\i}lek} M.,  {Thies} I.,  {Kroupa} P.,   {Famaey} B.,  2018, \mn@doi
  [\aap] {10.1051/0004-6361/201731939}, \href
  {https://ui.adsabs.harvard.edu/abs/2018A\&A...614A..59B} {614, A59}

\bibitem[\protect\citeauthoryear{{Calzetti}}{{Calzetti}}{2008}]{Calzetti08}
{Calzetti} D.,  2008, in {Knapen} J.~H.,  {Mahoney} T.~J.,   {Vazdekis} A.,
  eds,  Astronomical Society of the Pacific Conference Series Vol. 390,
  Pathways Through an Eclectic Universe. p.~121 (\mn@eprint {arXiv}
  {0707.0467})

\bibitem[\protect\citeauthoryear{{Calzetti}}{{Calzetti}}{2013}]{Calzetti13}
{Calzetti} D.,  2013, {Star Formation Rate Indicators}.
p.~419

\bibitem[\protect\citeauthoryear{{Dabringhausen}}{{Dabringhausen}}{2019}]{Dabring19}
{Dabringhausen} J.,  2019, \mn@doi [\mnras] {10.1093/mnras/stz2562}, \href
  {https://ui.adsabs.harvard.edu/abs/2019MNRAS.490..848D} {490, 848}

\bibitem[\protect\citeauthoryear{{Dav{\'e}}}{{Dav{\'e}}}{2008}]{Dave08}
{Dav{\'e}} R.,  2008, \mn@doi [\mnras] {10.1111/j.1365-2966.2008.12866.x},
  \href {https://ui.adsabs.harvard.edu/abs/2008MNRAS.385..147D} {385, 147}

\bibitem[\protect\citeauthoryear{{De Propris} et~al.,}{{De Propris}
  et~al.}{2014}]{DePropris+14}
{De Propris} R.,  et~al., 2014, \mn@doi [\mnras] {10.1093/mnras/stu1452}, \href
  {https://ui.adsabs.harvard.edu/abs/2014MNRAS.444.2200D} {444, 2200}

\bibitem[\protect\citeauthoryear{{Delgado-Serrano}, {Hammer}, {Yang}, {Puech},
  {Flores}  \& {Rodrigues}}{{Delgado-Serrano}
  et~al.}{2010}]{Delgado-Serrano+10}
{Delgado-Serrano} R.,  {Hammer} F.,  {Yang} Y.~B.,  {Puech} M.,  {Flores} H.,
  {Rodrigues} M.,  2010, \mn@doi [\aap] {10.1051/0004-6361/200912704}, \href
  {https://ui.adsabs.harvard.edu/abs/2010A\&A...509A..78D} {509, A78}

\bibitem[\protect\citeauthoryear{{Disney}, {Romano}, {Garcia-Appadoo}, {West},
  {Dalcanton}  \& {Cortese}}{{Disney} et~al.}{2008}]{Disney+08}
{Disney} M.~J.,  {Romano} J.~D.,  {Garcia-Appadoo} D.~A.,  {West} A.~A.,
  {Dalcanton} J.~J.,   {Cortese} L.,  2008, \mn@doi [\nat]
  {10.1038/nature07366}, \href
  {https://ui.adsabs.harvard.edu/abs/2008Natur.455.1082D} {455, 1082}

\bibitem[\protect\citeauthoryear{{Elmegreen}}{{Elmegreen}}{1997}]{Elmegreen97}
{Elmegreen} B.~G.,  1997, \mn@doi [\apj] {10.1086/304562}, \href
  {https://ui.adsabs.harvard.edu/abs/1997ApJ...486..944E} {486, 944}

\bibitem[\protect\citeauthoryear{{Elmegreen}}{{Elmegreen}}{1999}]{Elmegreen99}
{Elmegreen} B.~G.,  1999, \mn@doi [\apj] {10.1086/307011}, \href
  {https://ui.adsabs.harvard.edu/abs/1999ApJ...515..323E} {515, 323}

\bibitem[\protect\citeauthoryear{{Famaey} \& {McGaugh}}{{Famaey} \&
  {McGaugh}}{2012}]{FamaeyMcGaugh12}
{Famaey} B.,  {McGaugh} S.~S.,  2012, \mn@doi [Living Reviews in Relativity]
  {10.12942/lrr-2012-10}, \href
  {https://ui.adsabs.harvard.edu/abs/2012LRR....15...10F} {15, 10}

\bibitem[\protect\citeauthoryear{{Grebel}}{{Grebel}}{1997}]{Grebel97}
{Grebel} E.~K.,  1997, Reviews in Modern Astronomy, \href
  {https://ui.adsabs.harvard.edu/abs/1997RvMA...10...29G} {10, 29}

\bibitem[\protect\citeauthoryear{{Grebel}, {Gallagher}  \& {Harbeck}}{{Grebel}
  et~al.}{2003}]{Grebel+03}
{Grebel} E.~K.,  {Gallagher} John~S. I.,   {Harbeck} D.,  2003, \mn@doi [\aj]
  {10.1086/368363}, \href
  {https://ui.adsabs.harvard.edu/abs/2003AJ....125.1926G} {125, 1926}

\bibitem[\protect\citeauthoryear{{Gunawardhana} et~al.,}{{Gunawardhana}
  et~al.}{2011}]{Gunawardhana+11}
{Gunawardhana} M.~L.~P.,  et~al., 2011, \mn@doi [\mnras]
  {10.1111/j.1365-2966.2011.18800.x}, \href
  {https://ui.adsabs.harvard.edu/abs/2011MNRAS.415.1647G} {415, 1647}

\bibitem[\protect\citeauthoryear{{Haghi}, {Amiri}, {Hasani Zonoozi}, {Banik},
  {Kroupa}  \& {Haslbauer}}{{Haghi} et~al.}{2019}]{Haghi+19}
{Haghi} H.,  {Amiri} V.,  {Hasani Zonoozi} A.,  {Banik} I.,  {Kroupa} P.,
  {Haslbauer} M.,  2019, \mn@doi [\apjl] {10.3847/2041-8213/ab4517}, \href
  {https://ui.adsabs.harvard.edu/abs/2019ApJ...884L..25H} {884, L25}

\bibitem[\protect\citeauthoryear{{Hopkins}}{{Hopkins}}{2018}]{Hopkins18}
{Hopkins} A.~M.,  2018, \mn@doi [\pasa] {10.1017/pasa.2018.29}, \href
  {https://ui.adsabs.harvard.edu/abs/2018PASA...35...39H} {35, 39}

\bibitem[\protect\citeauthoryear{{Hoversten} \& {Glazebrook}}{{Hoversten} \&
  {Glazebrook}}{2008}]{HG08}
{Hoversten} E.~A.,  {Glazebrook} K.,  2008, \mn@doi [\apj] {10.1086/524095},
  \href {https://ui.adsabs.harvard.edu/abs/2008ApJ...675..163H} {675, 163}

\bibitem[\protect\citeauthoryear{{Ibata} et~al.,}{{Ibata}
  et~al.}{2013}]{Ibata_2013}
{Ibata} R.~A.,  et~al., 2013, \mn@doi [Nature] {10.1038/nature11717}, \href
  {http://adsabs.harvard.edu/abs/2013Natur.493...62I} {493, 62}

\bibitem[\protect\citeauthoryear{{Ilbert} et~al.,}{{Ilbert}
  et~al.}{2015}]{Ilbert+15}
{Ilbert} O.,  et~al., 2015, \mn@doi [\aap] {10.1051/0004-6361/201425176}, \href
  {https://ui.adsabs.harvard.edu/abs/2015A\&A...579A...2I} {579, A2}

\bibitem[\protect\citeauthoryear{{Javadi}, {van Loon}, {Khosroshahi},
  {Tabatabaei}, {Hamedani Golshan}  \& {Rashidi}}{{Javadi}
  et~al.}{2017}]{Javadi+17}
{Javadi} A.,  {van Loon} J.~T.,  {Khosroshahi} H.~G.,  {Tabatabaei} F.,
  {Hamedani Golshan} R.,   {Rashidi} M.,  2017, \mn@doi [\mnras]
  {10.1093/mnras/stw2463}, \href
  {https://ui.adsabs.harvard.edu/abs/2017MNRAS.464.2103J} {464, 2103}

\bibitem[\protect\citeauthoryear{{Je{\v{r}}{\'a}bkov{\'a}}, {Hasani Zonoozi},
  {Kroupa}, {Beccari}, {Yan}, {Vazdekis}  \& {Zhang}}{{Je{\v{r}}{\'a}bkov{\'a}}
  et~al.}{2018}]{Jerabkova+18}
{Je{\v{r}}{\'a}bkov{\'a}} T.,  {Hasani Zonoozi} A.,  {Kroupa} P.,  {Beccari}
  G.,  {Yan} Z.,  {Vazdekis} A.,   {Zhang} Z.~Y.,  2018, \mn@doi [\aap]
  {10.1051/0004-6361/201833055}, \href
  {https://ui.adsabs.harvard.edu/abs/2018A\&A...620A..39J} {620, A39}

\bibitem[\protect\citeauthoryear{{Karachentsev}}{{Karachentsev}}{2012}]{Karachentsev12}
{Karachentsev} I.~D.,  2012, \mn@doi [Astrophysical Bulletin]
  {10.1134/S1990341312020010}, \href
  {https://ui.adsabs.harvard.edu/abs/2012AstBu..67..123K} {67, 123}

\bibitem[\protect\citeauthoryear{{Karachentsev} \& {Kaisina}}{{Karachentsev} \&
  {Kaisina}}{2013}]{KK13}
{Karachentsev} I.~D.,  {Kaisina} E.~I.,  2013, \mn@doi [\aj]
  {10.1088/0004-6256/146/3/46}, \href
  {https://ui.adsabs.harvard.edu/abs/2013AJ....146...46K} {146, 46}

\bibitem[\protect\citeauthoryear{{Karachentsev}, {Karachentseva}, {Huchtmeier}
  \& {Makarov}}{{Karachentsev} et~al.}{2004}]{Karachentsev+04}
{Karachentsev} I.~D.,  {Karachentseva} V.~E.,  {Huchtmeier} W.~K.,   {Makarov}
  D.~I.,  2004, \mn@doi [\aj] {10.1086/382905}, \href
  {https://ui.adsabs.harvard.edu/abs/2004AJ....127.2031K} {127, 2031}

\bibitem[\protect\citeauthoryear{{Karachentsev}, {Makarov}  \&
  {Kaisina}}{{Karachentsev} et~al.}{2013}]{Karachentsev+13}
{Karachentsev} I.~D.,  {Makarov} D.~I.,   {Kaisina} E.~I.,  2013, \mn@doi [\aj]
  {10.1088/0004-6256/145/4/101}, \href
  {https://ui.adsabs.harvard.edu/abs/2013AJ....145..101K} {145, 101}

\bibitem[\protect\citeauthoryear{{Keenan}, {Barger}  \& {Cowie}}{{Keenan}
  et~al.}{2013}]{Keenan+13}
{Keenan} R.~C.,  {Barger} A.~J.,   {Cowie} L.~L.,  2013, \mn@doi [\apj]
  {10.1088/0004-637X/775/1/62}, \href
  {https://ui.adsabs.harvard.edu/abs/2013ApJ...775...62K} {775, 62}

\bibitem[\protect\citeauthoryear{{Kennicutt} \& {Evans}}{{Kennicutt} \&
  {Evans}}{2012}]{KE12}
{Kennicutt} R.~C.,  {Evans} N.~J.,  2012, \mn@doi [\araa]
  {10.1146/annurev-astro-081811-125610}, \href
  {https://ui.adsabs.harvard.edu/abs/2012ARA&A..50..531K} {50, 531}

\bibitem[\protect\citeauthoryear{{Kere{\v{s}}}, {Katz}, {Weinberg}  \&
  {Dav{\'e}}}{{Kere{\v{s}}} et~al.}{2005}]{Keres+05}
{Kere{\v{s}}} D.,  {Katz} N.,  {Weinberg} D.~H.,   {Dav{\'e}} R.,  2005,
  \mn@doi [\mnras] {10.1111/j.1365-2966.2005.09451.x}, \href
  {https://ui.adsabs.harvard.edu/abs/2005MNRAS.363....2K} {363, 2}

\bibitem[\protect\citeauthoryear{{Knox}, {Hawkins}  \& {Hambly}}{{Knox}
  et~al.}{1999}]{Knox_1999}
{Knox} R.~A.,  {Hawkins} M.~R.~S.,   {Hambly} N.~C.,  1999, \mn@doi [MNRAS]
  {10.1046/j.1365-8711.1999.02625.x}, \href
  {https://ui.adsabs.harvard.edu/abs/1999MNRAS.306..736K} {306, 736}

\bibitem[\protect\citeauthoryear{{Kormendy}, {Drory}, {Bender}  \&
  {Cornell}}{{Kormendy} et~al.}{2010}]{Kormendy+10}
{Kormendy} J.,  {Drory} N.,  {Bender} R.,   {Cornell} M.~E.,  2010, \mn@doi
  [\apj] {10.1088/0004-637X/723/1/54}, \href
  {https://ui.adsabs.harvard.edu/abs/2010ApJ...723...54K} {723, 54}

\bibitem[\protect\citeauthoryear{{Kroupa}}{{Kroupa}}{2002}]{Kroupa02}
{Kroupa} P.,  2002, \mn@doi [\mnras] {10.1046/j.1365-8711.2002.05128.x}, \href
  {https://ui.adsabs.harvard.edu/abs/2002MNRAS.330..707K} {330, 707}

\bibitem[\protect\citeauthoryear{{Kroupa}}{{Kroupa}}{2015}]{Kroupa2015}
{Kroupa} P.,  2015, \mn@doi [Canadian Journal of Physics]
  {10.1139/cjp-2014-0179}, \href
  {https://ui.adsabs.harvard.edu/abs/2015CaJPh..93..169K} {93, 169}

\bibitem[\protect\citeauthoryear{{Kroupa}, {Theis}  \& {Boily}}{{Kroupa}
  et~al.}{2005}]{Kroupa+05}
{Kroupa} P.,  {Theis} C.,   {Boily} C.~M.,  2005, \mn@doi [\aap]
  {10.1051/0004-6361:20041122}, \href
  {https://ui.adsabs.harvard.edu/abs/2005A\&A...431..517K} {431, 517}

\bibitem[\protect\citeauthoryear{{Kroupa}, {Weidner}, {Pflamm-Altenburg},
  {Thies}, {Dabringhausen}, {Marks}  \& {Maschberger}}{{Kroupa}
  et~al.}{2013}]{Kroupa+13}
{Kroupa} P.,  {Weidner} C.,  {Pflamm-Altenburg} J.,  {Thies} I.,
  {Dabringhausen} J.,  {Marks} M.,   {Maschberger} T.,  2013, {The Stellar and
  Sub-Stellar Initial Mass Function of Simple and Composite Populations}.
p.~115, \mn@doi{10.1007/978-94-007-5612-0_4}

\bibitem[\protect\citeauthoryear{{Lee} et~al.,}{{Lee} et~al.}{2009}]{Lee+09}
{Lee} J.~C.,  et~al., 2009, \mn@doi [\apj] {10.1088/0004-637X/706/1/599}, \href
  {https://ui.adsabs.harvard.edu/abs/2009ApJ...706..599L} {706, 599}

\bibitem[\protect\citeauthoryear{{Lelli}, {McGaugh}  \& {Schombert}}{{Lelli}
  et~al.}{2016}]{Lelli+16}
{Lelli} F.,  {McGaugh} S.~S.,   {Schombert} J.~M.,  2016, \mn@doi [\aj]
  {10.3847/0004-6256/152/6/157}, \href
  {https://ui.adsabs.harvard.edu/abs/2016AJ....152..157L} {152, 157}

\bibitem[\protect\citeauthoryear{{Lelli}, {McGaugh}, {Schombert}  \&
  {Pawlowski}}{{Lelli} et~al.}{2017}]{Lelli+17}
{Lelli} F.,  {McGaugh} S.~S.,  {Schombert} J.~M.,   {Pawlowski} M.~S.,  2017,
  \mn@doi [\apj] {10.3847/1538-4357/836/2/152}, \href
  {https://ui.adsabs.harvard.edu/abs/2017ApJ...836..152L} {836, 152}

\bibitem[\protect\citeauthoryear{{McGaugh}}{{McGaugh}}{2012}]{McGaugh12}
{McGaugh} S.~S.,  2012, \mn@doi [\aj] {10.1088/0004-6256/143/2/40}, \href
  {https://ui.adsabs.harvard.edu/abs/2012AJ....143...40M} {143, 40}

\bibitem[\protect\citeauthoryear{{Meschin}, {Gallart}, {Aparicio}, {Hidalgo},
  {Monelli}, {Stetson}  \& {Carrera}}{{Meschin} et~al.}{2014}]{Meschin+14}
{Meschin} I.,  {Gallart} C.,  {Aparicio} A.,  {Hidalgo} S.~L.,  {Monelli} M.,
  {Stetson} P.~B.,   {Carrera} R.,  2014, \mn@doi [\mnras]
  {10.1093/mnras/stt2220}, \href
  {https://ui.adsabs.harvard.edu/abs/2014MNRAS.438.1067M} {438, 1067}

\bibitem[\protect\citeauthoryear{{Milgrom}}{{Milgrom}}{1983}]{Milgrom_1983}
{Milgrom} M.,  1983, \mn@doi [ApJ] {10.1086/161130}, \href
  {http://adsabs.harvard.edu/abs/1983ApJ...270..365M} {270, 365}

\bibitem[\protect\citeauthoryear{{Mor}, {Robin}, {Figueras}, {Roca-F{\`a}brega}
   \& {Luri}}{{Mor} et~al.}{2019}]{Mor+19}
{Mor} R.,  {Robin} A.~C.,  {Figueras} F.,  {Roca-F{\`a}brega} S.,   {Luri} X.,
  2019, \mn@doi [\aap] {10.1051/0004-6361/201935105}, \href
  {https://ui.adsabs.harvard.edu/abs/2019A\&A...624L...1M} {624, L1}

\bibitem[\protect\citeauthoryear{{Oehm}, {Thies}  \& {Kroupa}}{{Oehm}
  et~al.}{2017}]{Oehm+17}
{Oehm} W.,  {Thies} I.,   {Kroupa} P.,  2017, \mn@doi [\mnras]
  {10.1093/mnras/stw3381}, \href
  {https://ui.adsabs.harvard.edu/abs/2017MNRAS.467..273O} {467, 273}

\bibitem[\protect\citeauthoryear{{Patel}, {Besla}  \& {Sohn}}{{Patel}
  et~al.}{2017}]{Patel+17}
{Patel} E.,  {Besla} G.,   {Sohn} S.~T.,  2017, \mn@doi [\mnras]
  {10.1093/mnras/stw2616}, \href
  {https://ui.adsabs.harvard.edu/abs/2017MNRAS.464.3825P} {464, 3825}

\bibitem[\protect\citeauthoryear{{Pawlowski} \& {Kroupa}}{{Pawlowski} \&
  {Kroupa}}{2020}]{PK20}
{Pawlowski} M.~S.,  {Kroupa} P.,  2020, \mn@doi [\mnras]
  {10.1093/mnras/stz3163}, \href
  {https://ui.adsabs.harvard.edu/abs/2020MNRAS.491.3042P} {491, 3042}

\bibitem[\protect\citeauthoryear{{Pawlowski}, {Kroupa}  \&
  {Jerjen}}{{Pawlowski} et~al.}{2013}]{Pawlowski+13}
{Pawlowski} M.~S.,  {Kroupa} P.,   {Jerjen} H.,  2013, \mn@doi [\mnras]
  {10.1093/mnras/stt1384}, \href
  {https://ui.adsabs.harvard.edu/abs/2013MNRAS.435.1928P} {435, 1928}

\bibitem[\protect\citeauthoryear{{Peebles} \& {Nusser}}{{Peebles} \&
  {Nusser}}{2010}]{PeeblesNusser10}
{Peebles} P.~J.~E.,  {Nusser} A.,  2010, \mn@doi [\nat] {10.1038/nature09101},
  \href {https://ui.adsabs.harvard.edu/abs/2010Natur.465..565P} {465, 565}

\bibitem[\protect\citeauthoryear{{Pflamm-Altenburg} \&
  {Kroupa}}{{Pflamm-Altenburg} \& {Kroupa}}{2009}]{PK09}
{Pflamm-Altenburg} J.,  {Kroupa} P.,  2009, \mn@doi [\apj]
  {10.1088/0004-637X/706/1/516}, \href
  {https://ui.adsabs.harvard.edu/abs/2009ApJ...706..516P} {706, 516}

\bibitem[\protect\citeauthoryear{{Pflamm-Altenburg}, {Weidner}  \&
  {Kroupa}}{{Pflamm-Altenburg} et~al.}{2009}]{PWK09}
{Pflamm-Altenburg} J.,  {Weidner} C.,   {Kroupa} P.,  2009, \mn@doi [\mnras]
  {10.1111/j.1365-2966.2009.14522.x}, \href
  {https://ui.adsabs.harvard.edu/abs/2009MNRAS.395..394P} {395, 394}

\bibitem[\protect\citeauthoryear{{Planck Collaboration XIII}}{{Planck
  Collaboration XIII}}{2016}]{Planck_2015}
{Planck Collaboration XIII} 2016, \mn@doi [A\&A] {10.1051/0004-6361/201525830},
  \href {http://adsabs.harvard.edu/abs/2016A\%26A...594A..13P} {594, A13}

\bibitem[\protect\citeauthoryear{{Popesso} et~al.,}{{Popesso}
  et~al.}{2019}]{Popesso+19}
{Popesso} P.,  et~al., 2019, \mn@doi [\mnras] {10.1093/mnras/sty3210}, \href
  {https://ui.adsabs.harvard.edu/abs/2019MNRAS.483.3213P} {483, 3213}

\bibitem[\protect\citeauthoryear{Renzini \& Peng}{Renzini \&
  Peng}{2015}]{Renzini_2015}
Renzini A.,  Peng Y.-J.,  2015, \mn@doi [The Astrophysical Journal]
  {10.1088/2041-8205/801/2/l29}, 801, L29

\bibitem[\protect\citeauthoryear{{Ruiz-Lara}, {Gallart}, {Bernard}  \&
  {Cassisi}}{{Ruiz-Lara} et~al.}{2020}]{Ruiz-Lara+20}
{Ruiz-Lara} T.,  {Gallart} C.,  {Bernard} E.~J.,   {Cassisi} S.,  2020,
  preprint, \href {https://ui.adsabs.harvard.edu/abs/2020arXiv200312577R}
  {Arxiv} (\mn@eprint {arXiv} {2003.12577})

\bibitem[\protect\citeauthoryear{Shankar et~al.,}{Shankar
  et~al.}{2014}]{Shankar+14}
Shankar F.,  et~al., 2014, \mn@doi [Monthly Notices of the Royal Astronomical
  Society] {10.1093/mnras/stt2470}, 439, 3189

\bibitem[\protect\citeauthoryear{{Speagle}, {Steinhardt}, {Capak}  \&
  {Silverman}}{{Speagle} et~al.}{2014}]{Speagle+14}
{Speagle} J.~S.,  {Steinhardt} C.~L.,  {Capak} P.~L.,   {Silverman} J.~D.,
  2014, \mn@doi [\apjs] {10.1088/0067-0049/214/2/15}, \href
  {https://ui.adsabs.harvard.edu/abs/2014ApJS..214...15S} {214, 15}

\bibitem[\protect\citeauthoryear{{Steinhardt}, {Capak}, {Masters}  \&
  {Speagle}}{{Steinhardt} et~al.}{2016}]{Steinhardt+16}
{Steinhardt} C.~L.,  {Capak} P.,  {Masters} D.,   {Speagle} J.~S.,  2016,
  \mn@doi [\apj] {10.3847/0004-637X/824/1/21}, \href
  {http://adsabs.harvard.edu/abs/2016ApJ...824...21S} {824, 21}

\bibitem[\protect\citeauthoryear{{Watts}, {Meurer}, {Lagos}, {Bruzzese},
  {Kroupa}  \& {Jerabkova}}{{Watts} et~al.}{2018}]{Watts+18}
{Watts} A.~B.,  {Meurer} G.~R.,  {Lagos} C. D.~P.,  {Bruzzese} S.~M.,  {Kroupa}
  P.,   {Jerabkova} T.,  2018, \mn@doi [\mnras] {10.1093/mnras/sty1006}, \href
  {https://ui.adsabs.harvard.edu/abs/2018MNRAS.477.5554W} {477, 5554}

\bibitem[\protect\citeauthoryear{Williams et~al.,}{Williams
  et~al.}{2017}]{Williams+17}
Williams B.~F.,  et~al., 2017, \mn@doi [The Astrophysical Journal]
  {10.3847/1538-4357/aa862a}, 846, 145

\bibitem[\protect\citeauthoryear{{Wittenburg}, {Kroupa}  \&
  {Famaey}}{{Wittenburg} et~al.}{2020}]{Wittenburg+20}
{Wittenburg} N.,  {Kroupa} P.,   {Famaey} B.,  2020, \mn@doi [\apj]
  {10.3847/1538-4357/ab6d73}, \href
  {https://ui.adsabs.harvard.edu/abs/2020ApJ...890..173W} {890, 173}

\bibitem[\protect\citeauthoryear{{Yan}, {Jerabkova}  \& {Kroupa}}{{Yan}
  et~al.}{2017}]{Yan+17}
{Yan} Z.,  {Jerabkova} T.,   {Kroupa} P.,  2017, \mn@doi [\aap]
  {10.1051/0004-6361/201730987}, \href
  {https://ui.adsabs.harvard.edu/abs/2017A\&A...607A.126Y} {607, A126}

\bibitem[\protect\citeauthoryear{{Yan}, {Jerabkova}, {Kroupa}  \&
  {Vazdekis}}{{Yan} et~al.}{2019}]{Yan+19}
{Yan} Z.,  {Jerabkova} T.,  {Kroupa} P.,   {Vazdekis} A.,  2019, \mn@doi [\aap]
  {10.1051/0004-6361/201936029}, \href
  {https://ui.adsabs.harvard.edu/abs/2019A\&A...629A..93Y} {629, A93}

\bibitem[\protect\citeauthoryear{{Yan}, {Jerabkova}  \& {Kroupa}}{{Yan}
  et~al.}{2020}]{Yan+20}
{Yan} Z.,  {Jerabkova} T.,   {Kroupa} P.,  2020, \mn@doi [A\&A]
  {10.1051/0004-6361/202037567}, \href
  {https://ui.adsabs.harvard.edu/abs/2020A\&A...637A..68Y} {637, A68}

\bibitem[\protect\citeauthoryear{{Zhao}, {Famaey}, {L{\"u}ghausen}  \&
  {Kroupa}}{{Zhao} et~al.}{2013}]{Zhao+13}
{Zhao} H.,  {Famaey} B.,  {L{\"u}ghausen} F.,   {Kroupa} P.,  2013, \mn@doi
  [\aap] {10.1051/0004-6361/201321879}, \href
  {https://ui.adsabs.harvard.edu/abs/2013A\&A...557L...3Z} {557, L3}

\bibitem[\protect\citeauthoryear{{Zonoozi}, {Mahani}  \& {Kroupa}}{{Zonoozi}
  et~al.}{2019}]{Zonoozi+19}
{Zonoozi} A.~H.,  {Mahani} H.,   {Kroupa} P.,  2019, \mn@doi [\mnras]
  {10.1093/mnras/sty2812}, \href
  {https://ui.adsabs.harvard.edu/abs/2019MNRAS.483...46Z} {483, 46}

\bibitem[\protect\citeauthoryear{{da~Silva}, {Fumagalli}  \&
  {Krumholz}}{{da~Silva} et~al.}{2014}]{daSilva+14}
{da~Silva} R.~L.,  {Fumagalli} M.,   {Krumholz} M.~R.,  2014, \mn@doi [\mnras]
  {10.1093/mnras/stu1688}, \href
  {https://ui.adsabs.harvard.edu/abs/2014MNRAS.444.3275D} {444, 3275}

\makeatother
\end{thebibliography}

\bsp
\label{lastpage} 
\end{document}